\documentclass[aps,prl,twocolumn,superscriptaddress,showpacs]{revtex4}

\usepackage{graphicx}
\usepackage{bm}

\begin{document}

\newcommand{\np}{\mbox{$ n p \rightarrow d\pi^0$ }}
\newcommand{\npel}{\mbox{$ n p \rightarrow p n$ }}
\newcommand{\Afb}{\mbox{$A_{\rm fb}$}}
\newcommand{\Afbtheta}{\mbox{$A_{\rm fb}(\theta )$}}
\newcommand{\LH}{\mbox{$\rm LH_2$}}
\newcommand{\GEANT}{\textsc{GEANT}}
\newcommand{\TRIUMF}{\textsc{TRIUMF}}
\newcommand{\SASP}{\textsc{SASP}}

%\preprint{}

%\title{Charge symmetry breaking in $\mathbf{n p \rightarrow d \pi^0}$ }
\title{Charge symmetry breaking in $ \bm {n p \rightarrow d \pi^0}$ }


\author{A.K. Opper}
\email[]{opper@ohiou.edu}
%\thanks{}
\affiliation{Ohio University, Athens, Ohio 45701}
\author{E. Korkmaz}
\affiliation{University of Northern British Columbia, 
             Prince George, British Columbia, Canada}
\author{D.A. Hutcheon}
\affiliation{University of Alberta, Edmonton, Alberta, Canada }
\affiliation{TRIUMF, Vancouver, British Columbia, Canada}
\author{R. Abegg}
\altaffiliation{deceased}
\affiliation{University of Alberta, Edmonton, Alberta, Canada }
\affiliation{TRIUMF, Vancouver, British Columbia, Canada}
\author{C.A. Davis}
\affiliation{TRIUMF, Vancouver, British Columbia, Canada}
\affiliation{University of Manitoba, Winnipeg, Manitoba, Canada}
\author{R.W. Finlay}
\affiliation{Ohio University, Athens, Ohio 45701}
\author{P.W. Green}
\affiliation{University of Alberta, Edmonton, Alberta, Canada }
\affiliation{TRIUMF, Vancouver, British Columbia, Canada}
\author{L.G. Greeniaus}
\affiliation{University of Alberta, Edmonton, Alberta, Canada }
\affiliation{TRIUMF, Vancouver, British Columbia, Canada}
\author{D.V. Jordan}
\altaffiliation[Current address:  ]{Pacific Northwest National Labs, Richland,
               Washington, 99352}
\affiliation{Ohio University, Athens, Ohio 45701}
\affiliation{University of Alberta, Edmonton, Alberta, Canada }
\author{J.A. Niskanen}
\affiliation{University of Helsinki, Helsinki, Finland} 
\author{G.V. O'Rielly}
\altaffiliation[Current address: ]{University of Massachussets, Dartmouth, 
   North Dartmouth, MA \ \ 02747}
\affiliation{University of Northern British Columbia, 
            Prince George, British Columbia, Canada}
\author{T.A. Porcelli}
\affiliation{University of Northern British Columbia, 
           Prince George, British Columbia, Canada}
\author{S.D. Reitzner}
\altaffiliation[Current address: ]{The Ohio State University, Columbus,
                Ohio, 43210}
\affiliation{Ohio University, Athens, Ohio 45701}
\author{P.L. Walden}
\affiliation{TRIUMF, Vancouver, British Columbia, Canada}
\affiliation{University of British Columbia, 
         Vancouver, British Columbia, Canada}
\author{S. Yen}
\affiliation{TRIUMF, Vancouver, British Columbia, Canada}


\date{\today}

\begin{abstract}
The forward--backward asymmetry in \np,  which must be zero in the
center-of-mass system if charge symmetry is respected, has been measured
to be $[17.2 \pm 8 {\rm (stat)} \pm 5.5 {\rm (sys)}] \times 10^{-4}$,
at an incident neutron energy of 279.5 MeV.  This charge symmetry
breaking observable was extracted by fitting the data with GEANT-based
simulations
and is compared to recent chiral effective field theory
calculations, with implications regarding the value of the $u \ d$ quark 
mass difference.
\end{abstract}

%\keywords{}

\maketitle


In the quark model, the breaking of charge independence and charge 
symmetry arises from the mass difference of the $up$ and $down$ current 
quarks and the electromagnetic interaction between quarks.
The basic $np$ interaction is particularly sensitive
to such fundamental effects since the ``background" Coulomb 
force is absent in this system. Indeed, charge symmetry breaking (CSB) has been
unambiguously observed \cite{tri-csb,tri-new, iucf-csb} in $np$ elastic 
scattering
at three different energies. Measurement of CSB in the inelastic
\np reaction complements the existing data in that it is sensitive to
contributions that are absent in the elastic channel. Furthermore, 
this reaction is unique as a testing ground for effective field
theory calculations addressing the important issue of isospin symmetry 
violation in pion-nucleon scattering. The observable of interest in \np is 
the center-of-mass
forward--backward asymmetry, $A_{\rm fb}$, which we define as
\begin{equation}
  A_{\rm fb}(\theta ) \equiv
    { {\sigma (\theta ) - \sigma (\pi - \theta )} \over
      {\sigma (\theta ) + \sigma (\pi - \theta )} }
\label{eq:Afb-def-eq}
\end{equation}
where $\theta$ is the angle between the incident beam and the scattered
deuteron. 
Note that the asymmetry must be zero if charge symmetry is conserved. 
We report on a measurement of this asymmetry at a 
neutron energy a few MeV above the reaction threshold (275.06 MeV), and 
compare our result to recent theoretical predictions \cite{Ni99,vKNM} 
bearing on such fundamental questions as the $u \ d$ quark mass difference 
and our understanding of QCD dynamics and symmetries in low-energy hadronic 
interactions.

\section{The Experiment\label{expt-sec}}

\begin{center}
\begin{figure}[htbp]
\vspace{6.8cm}
\includegraphics{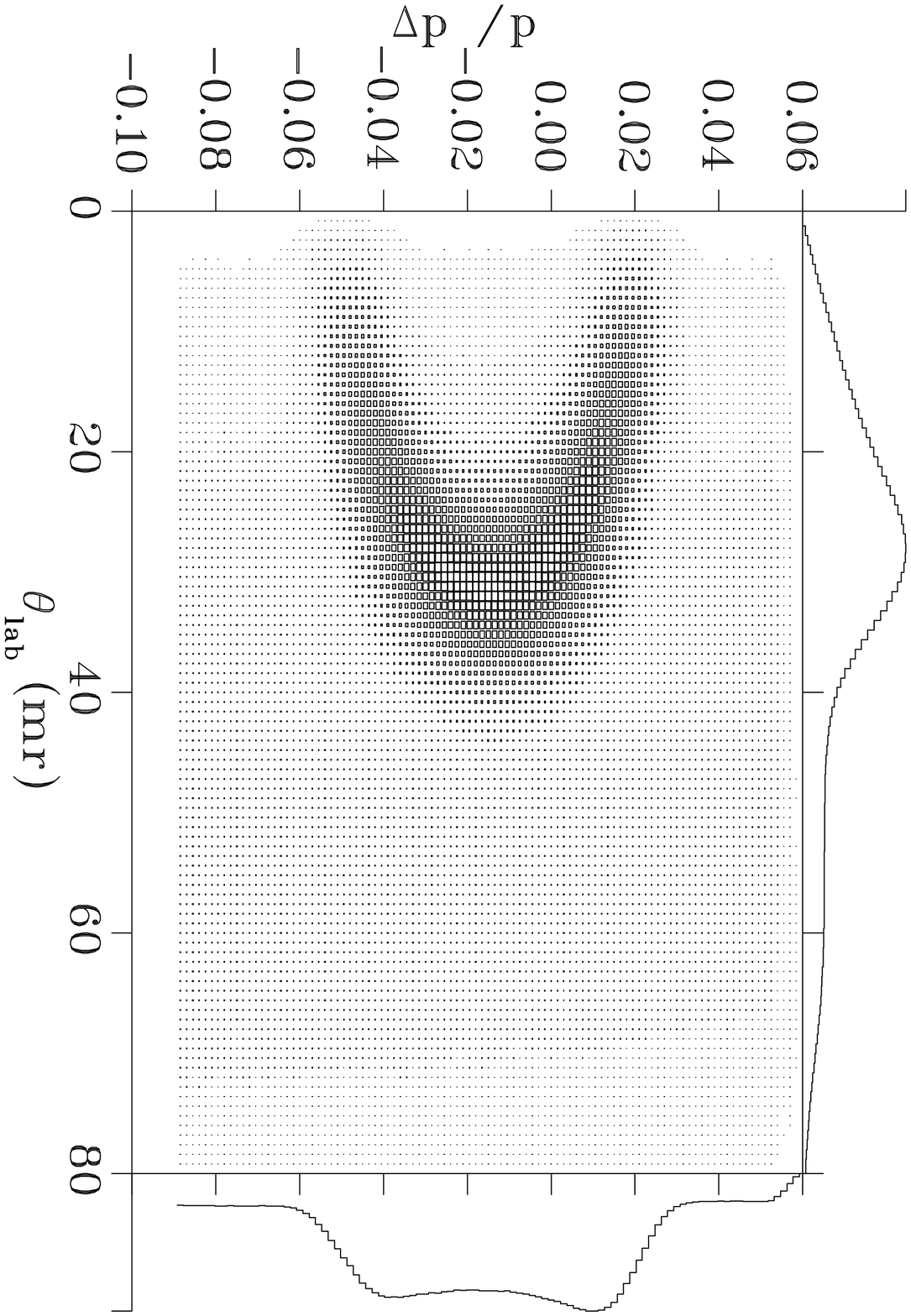}
\caption[]{Kinematic locus of  \np\ data. }
\label{locus-fig}
\end{figure}
\end{center}

The experiment was performed at TRIUMF with a 279.5 MeV neutron beam, a 
liquid hydrogen target, and the \SASP\ magnetic spectrometer \cite{Wa99} 
positioned at $0^{\circ}$. With these near threshold kinematics 
and the large acceptance of \SASP, 
the full deuteron distribution from \np was detected in one setting of the 
spectrometer thereby eliminating many systematic uncertainties. These 
deuterons form a distinct kinematic locus in momentum vs laboratory 
scattering angle, which is shown in fig.~\ref{locus-fig} for the 
collected data.
%%%%%%%%%%%%% Figure: np->dpi locus -- all data  %%%%%%%%%%%%%%%%%
%%%%%%%%%%%%%%%%%%%%%%%%%%%%%%%%%%%%%%%%%%%%%%%%%%%%%%%%%%%%%%%%% 

The \TRIUMF\ \textsc{CHARGEX} facility \cite{He87} produced
the neutron beam by passing a high intensity proton beam through a
 thin $\rm ^7Li$ target.  A sweeping magnet deflected the primary  proton 
beam into a well- shielded dump.  The liquid hydrogen target
($\rm LH_2$) was centered 92 cm downstream from the $\rm ^7Li$ target and was
contained within a flat cylindrical volume, 10 cm in diameter with a nominal
thickness of 2 cm.  Two sets of veto counters (FEV1, FEV2) and a  trigger 
counter set (FET) were each composed of a  pair of plastic scintillators
positioned above one another.  This allowed more stable operation in the 
high (few MHz) particle rate environment. 
The thick  veto scintillators were upstream
of the \LH\ and shadowed it.  The FET counters were positioned immediately
downstream of the \LH.

        Three multi-wire proportional chambers, positioned upstream of the
\SASP\ entrance (FECs i.e. Front-End Chambers), provided tracking
information for charged particles.  Each FEC consisted of a pair of orthogonal
wire planes.  The first and last FECs, separated by 33 cm, were mounted
to measure vertical and horizontal coordinates.  The third FEC was positioned
midway between the other two and rotated $40^\circ$ with respect to them
for efficiency measurements and to aid in multi-hit track reconstruction.
Particle tracking near the \SASP\ 
focal plane was provided by two vertical-drift
chambers (VDCs). 
Three sets of scintillators, downstream from  the VDCs,
provided timing and particle ID information as well as sufficient
redundancy to determine the efficiencies of all focal plane area detectors.

%%%%%%%%%%%%% Figure: np->dpi experimental layout  %%%%%%%%%%%%%%%%%
%\begin{center}
%\begin{figure}[htbp]
%\vspace{9.5cm}
%\special{psfile=layout.eps angle=00 hscale=50 vscale=50
%  hoffset=00 voffset=-50}
%\caption[]{Schematic of the \np experimental apparatus}
%\label{apparatus_fig}
%\end{figure}
%\end{center}
%%%%%%%%%%%%%%%%%%%%%%%%%%%%%%%%%%%%%%%%%%%%%%%%%%%%%%%%%%%%%%%%%

Measurements of $np$ elastic scattering with incident neutron beams
that filled the same target space and produced protons that spanned the
momentum distribution of the \np\ reaction provided a stringent test of the
description of the spectrometer acceptance.
Further details on the apparatus and other technical aspects of the
measurement are found in reference \cite{Hu01}.

\section{Extraction of  $\mathbf{A_{\rm {\bf fb} } }$ \label{extraction-sec}  }

  Close to threshold, the \np\ cross-section in the center-of-mass frame
is given by
\begin{equation}
\frac{d\sigma}{d\Omega} (\theta) = A_{0} +
A_{1}P_{1}(\cos\theta) + A_{2}P_{2}(\cos\theta),
\label{eq:x-section-eq}
\end{equation}
where $P_1$ and $P_2$ are Legendre polynomials. The $A_0$ and $A_2$
coefficients were previously measured \cite{Hu91} at a number of 
energies within 10 MeV above threshold. 
The presence of charge symmetry breaking is
reflected in the $A_{1}$ term as it is odd in $\cos\theta$. In this
standard parametrization, the angle integrated form of $A_{\rm fb}$ is
given by $A_{\rm fb} = \frac{1}{2}A_{1}/A_{0}$.

For a given beam energy, $\cos\theta$ varies linearly with the
longitudinal component of deuteron momentum in the laboratory
reference frame. Ideally, the  $\cos\theta$  distribution would be
found by a suitable, simple projection of the data of fig.~\ref{locus-fig}.
However, the measured deuteron locus is distorted by 
energy loss, multiple scattering, energy spread of the beam, and spectrometer
acceptance making a direct extraction of \Afb\ impossible.
Instead, the data were binned according to laboratory momentum and angle
(as in fig.~\ref{locus-fig})  and compared to a model which represented
the background due to $\rm C(n,d)$ reactions as a low-order polynomial
and generated the locus of $\rm H(n,d)\pi^0 $ events by Monte Carlo
simulation of the beam, target, reaction cross section,
spectrometer and detectors.

    The simulation was based on \textsc{GEANT3}.  It began with a proton beam
incident on the $\rm^7Li$ target and included energy loss by the proton
beam as well as the angular and energy distribution of neutrons
from the $\rm^7Li(p,n)$ reaction.  Production of deuterons according
to the distribution of equation~\ref{eq:x-section-eq}
was allowed in the \LH\ target and other hydrogenous material such as
scintillators and their wrapping.
Standard \GEANT\ tracking options were adopted for deuteron energy
loss and multiple scattering but the reaction losses, which amount
to 1--2\% and are momentum dependent, were parametrized from data
on deuteron elastic and reaction cross-sections from hydrogen and
carbon \cite{deut_cross}.  Tracking through the \SASP\ dipole used a field
map obtained at 875 Amp and scaled up to the operating current of 905 Amp.
Data were acquired in 10 different periods spanning  two years and the 
simulation accounted for  measured detector efficiencies, 
scintillator thresholds, missing FEC wires, and known changes in target 
thickness in a manner consistent with the actual running periods.

To reduce the possibility of psychological bias in matching simulation to
data, a blind analysis technique was used which incorporated a
hidden offset to the $A_1/A_0$ asymmetry parameter of the \np\ generator.
The collaborators developing the simulation and extracting the observable
did not know the value of the offset until all consistency checks had been
satisfied.

\section{Systematic Effects  \label{sys-sec}  }

\begin{center}
\begin{figure}[htbp]
\vspace{8.4cm}
\includegraphics{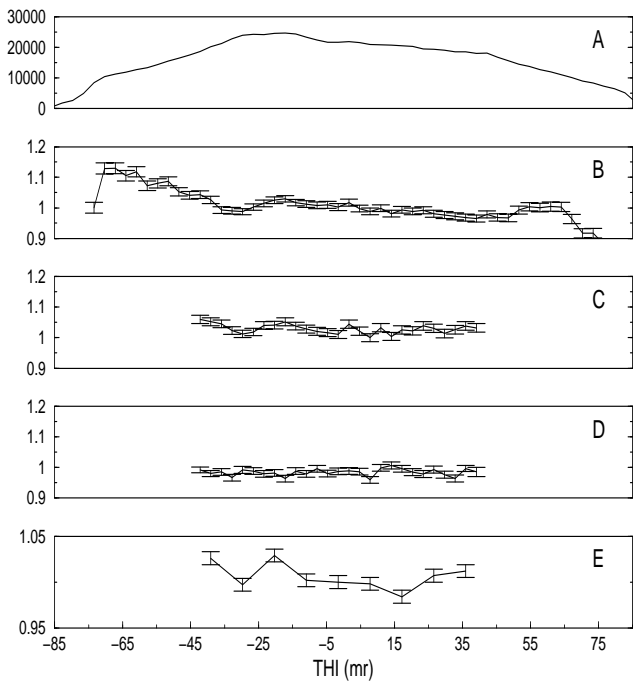}
\caption[]{The vertical projection of the lab scattering angle for the
center acceptance slice, elastic $np$ scattering;
A) data yield for $\delta $ = $-$4\%, no cuts;
B) data ratio ($-$4\%/+4\%), no cuts;
C) data ratio ($-$4\%/+4\%), full acceptance cuts;
D) simulation ratio ($-$4\%/+4\%), full acceptance cuts;
E) ratio of C to D, normalized to the center bin.}
\label{accep-fig}
\end{figure}
\end{center}

The acceptance of \SASP\ is a function of the initial target position and
direction of the deuteron as well as its momentum.
Non--uniformities in the momentum acceptance of \SASP\ would systematically
produce a false asymmetry and had to be limited.  
High-statistics data from $np$ elastic scattering were collected
and compared to model simulations to determine a fiducial volume of 
uniform acceptance.   
For these calibration  measurements, 
the \SASP\ magnets were set to  their values for the \np\ running, but the 
primary beam energy was adjusted so that the elastically scattered protons
had a momentum deviation $\delta = (p - p_0)/p_0 = -4, 0$ or $+4\% $ 
compared to the central  momentum of the deuterons of interest. 
Projections of the $np$ elastic data direction for position slices 
were formed, and the ratios of yields at $ -$4\% vs +4\% were formed for 
both data and simulation; see fig.~\ref{accep-fig}.
The analysis software acceptance cuts in position and direction were then 
limited to the regions common to both data and simulation which were 
uniform in momentum to the statistical precision of the data.

Simulation vs simulation comparisons were carried out to determine how
strongly experimental parameters and other effects  were correlated 
with $A_1/A_0$.  For example, momentum dependent deuteron reaction losses 
and detection
efficiencies are obvious mechanisms which can mimic the effect of a non--zero
$A_1/A_0$. 
Combining each correlation with the independently-determined uncertainty
of its parameter gave the systematic contributions shown in 
Table \ref{error_bud-tab}.   
However,
for the \LH\ target thickness, the proton beam energy ($T_{\rm beam}$) and the
central momentum of \SASP\ ($p_0$) the independent information was not
a sufficient constraint. 
Therefore, these three parameters, along with $A_1/A_0$, were treated 
as free parameters and their values extracted from fitting the data. 
To this end, simulations were made and  $\chi^2$ calculated for 81 
points in a four-dimensional space, in which each of the four free 
parameters was stepped above and below a nominal value.  
$\chi^2$ minimization techniques \cite{num_rec} were then used to obtain 
the values of the parameters at the global $\chi^2$ minimum, while the 
local curvature of the $\chi^2$ surface gave their errors and mutual 
correlations.

%%%%%%% Figure:  Acceptance Cuts (-4%/+4%) center Xi slice %%%%%%
%%%%%%%%%%%%%%%%%%%%%%%%%%%%%%%%%%%%%%%%%%%%%%%%%%%%%%%%%%%%%%%%%

\begin{table}%[H] add [H] placement to break table across pages
\caption{ Systematic Error Contributions to \Afb\ \label{error_bud-tab}}
\begin{ruledtabular}
\begin{tabular}{lc}
              & Uncert $ \times (10^{-4})$  \\  \hline
FEV threshold    & 2.5  \\
Separation between front and rear FECs        & 2.5  \\
Longitudinal position of $ \rm ^7Li$          & 2.5  \\
$A_2/A_0 $    & 2~~  \\
Deuteron reaction losses   & 1.5  \\
Detection efficiencies  & 1.5  \\
Primary beam energy spread    & 1~~  \\
Neutron angle & 1~~  \\
Background    & 1~~  \\
FET threshold    & 0.5 \\ \hline
\ \ \ Total  & 5.5 \\ %\hline
%              &     \\
%Fit \& Statistics & 8~~    \\ \hline
%              &      \\
%Total Uncert  & 10~~     \\
\end{tabular}
\end{ruledtabular}
\end{table}

\section{Results and Discussion  \label{result-sec}}

As a test of the model, the $\chi^2$ calculations and fitting were
repeated on subsets of the data and simulated data, selected
according to whether the reaction occurred in the Top or Bottom
part of the \LH\ target.  A second test divided events into those
originating in the Left or Right part of the target.
The best fit values and errors of $A_1/A_0$ (after removal of the
offset) and the other three parameters are presented in 
Table  \ref{params_subspaces-tab}.

The root mean square (rms) systematic error
for the full acceptance 
and the four subspaces is $\sim 2.7\%$ with the standard binning
scheme of 50 bins in $\delta$ and 20 bins in $\theta$, indicating 
a substantial discrepancy between data and the simulation.
Pixel by pixel examination of the contribution to $\chi^2$ revealed a
systematic difference in the profile of the locus along lines of steepest
ascent.  The sign of the differences tended to be positive at the peak and 
negative at both the ``inner'' and ``outer'' margins of the locus,
possibly due to inadequate treatment of deuteron scattering in the simulation.

A change in $A_1/A_0$ will not change the ratio of counts in peak vs
margins of the locus because it multiplies $\cos(\theta_{cm})$.  In contrast,
the \LH\ thickness, $p_0$, and $\rm T_{beam}$ all shift or broaden the 
locus and thus are sensitive to the ratio of locus counts at the peak vs 
margins.  
It is reasonable to expect further rebinning to remove sensitivity to 
unimportant details of the simulation without losing sensitivity to $A_1/A_0$.  
We repeated the $\chi^2$ grid search using 20 bins in $\delta$ and 10 bins in 
$\theta$, and again with 10 bins in $\delta$ and 5 bins in $\theta$.  
As expected, the fractional error dropped to 2.1\% and 1.4\%, respectively, 
with $A_1/A_0$ remaining consistent within errors.   
A more sophisticated binning scheme which treated the locus as a set of 
``elliptical'' and ``radial'' bins on top of rectangular background bins 
produced an rms error of 2.1\% for  2500 background bins and 36 locus bins, 
and an rms error of 1.2\% for 100 background bins and 6 locus bins.    
In all fits and binning schemes the best fit values of the asymmetry in 
the acceptance subspaces agreed within errors with the value for the full 
acceptance, which is $(34.4 \pm 16) \times 10^{-4}$, implying   
$A_{\rm fb} = [17.2 \pm 8 {\rm (stat)} \pm 5.5 {\rm (sys)}] \times 10^{-4}$.

\begin{table}%[H] add [H] placement to break table across pages
\caption{ Stability of the four free parameters over target subspaces;
  b = bottom; t = top; l = left; r = right.
 \label{params_subspaces-tab}}
\begin{ruledtabular}
\begin{tabular}{cclll}
  &  $(A_1/A_0)$  & relative  & relative   &  relative \\
  &  $(10^{-4})$  & $LH_2$ (mm) & $p_0$ (MeV/c) & $T_{\rm beam}$ (MeV) \\ 
                       \hline
full   & 34.4 $\pm$  16 & 0.94 $\pm$ 0.05 & 0.365 $\pm$ 0.015
                       & 0.048 $\pm$  0.001 \\
b & 30 $\pm$ 26  & 0.39 $\pm$ 0.09 & 0.547 $\pm$  0.025
                       & 0.086 $\pm$ 0.002 \\
t & 20  $\pm$ 20 & 1.14 $\pm$ 0.07 & 0.236 $\pm$  0.018
                       & 0.021 $\pm$  0.002 \\
l & 29 $\pm$ 23 & 1.21 $\pm$ 0.08 & 0.273 $\pm$  0.021
                     & 0.042 $ \pm$  0.002 \\
r & 15  $\pm$ 22 & 0.75 $\pm$ 0.08 & 0.427  $\pm$ 0.021
                 & 0.051  $\pm$ 0.002 \\
\end{tabular}
\end{ruledtabular}
\end{table}

Theoretical predictions of $A_{\rm fb}$ have been made  by Niskanen 
\cite{Ni99} using a meson-exchange coupled-channel model which showed
that the major contribution by far is due to $\pi \eta$ 
(and $\pi \eta'$) mixing in both the exchange and produced (outgoing) 
meson.
At our energy the prediction is $A_{\rm fb}= -28 \times 10^{-4}$, when 
accepted values are used for the $\eta NN$ coupling constant
and the $\pi^0 \eta$ mixing matrix element
($ g_{\eta NN}^2/4\pi = 3.68 $ from meson exchange $NN$ potential
models \cite{Du83}
and
$\langle \pi^0| {\cal H }| \eta \rangle = -0.0059$ $\rm GeV^2$ from
analysis of $\eta$ decay data \cite{Co86}).
More recently, $A_{\rm fb}$ was revisited \cite{vKNM} within the framework 
of chiral effective field theory where the issue of charge symmetery 
breaking in the rescattering amplitude of the exchanged pion was addressed.
The resulting additional contribution to \Afb\ is then expressed in terms 
of two parameters $\delta m_N$ and $\bar{\delta}m_N$ representing 
contributions from the $u \ d$ quark mass difference and from 
electromagnetic effects within the nucleon, respectively. 
Specifically, at our energy, \Afb\  is expressed as 
\begin{widetext}
  \begin{equation}
A_{\rm fb} = -0.28\% \times \left[ 
   \left( { {g_{\eta NN}} \over {\sqrt{4\pi (3.68)}}} \right)
     \left( { {\langle \pi^0| {\cal H }| \eta \rangle} \over
                          { -0.0059 {\rm \ GeV^2} } } \right)
     - { {0.87} \over {{\rm MeV} } } (\delta m_N -
        { {\bar{\delta}m_N} \over 2})  \right]
  \label{Afb-cont-eq}
  \end{equation}
\end{widetext}
where the first term arises from $\pi \eta $ mixing and the second from
$\pi^0 \  N $ scattering. With the introduction of the new term, 
\Afb\ changes sign and becomes positive with an estimated upper value around
$+69 \times 10^{-4}$, when large but reasonable values of 
$\delta m_N$ and $\bar{\delta}m_N$ are used \cite{vKNM}. 
Our positive experimental result strongly suggests, therefore, that such 
isospin violating $\pi^0 N$ interactions as outlined in reference \cite{vKNM}
are indeed significant. 

The parameters $\delta m_N$ and $\bar{\delta}m_N$ are also constrained by the
proton-neutron mass difference as
\begin{equation}
\Delta_N = m_n - m_p = \delta m_N + \bar{\delta}m_N = 1.29 \ {\rm MeV.}
\label{mn-mp-eq}
\end{equation}
When our \Afb\ result is combined with equations \ref{Afb-cont-eq} and 
\ref{mn-mp-eq}, and the values given above for the  $\eta NN$ coupling 
constant and $\pi \eta$ mixing matrix element, 
 we find that $\delta m_N = 1.66 \pm 0.27$ MeV and 
$\bar{\delta}m_N = -0.36 \pm 0.27$ MeV, assuming 
no theoretical uncertainties. 
We emphasize, however, that this last exercise is only meant to illustrate 
the significance and potential important implications of our \Afb\ result. 
Further theoretical studies are currently underway 
\cite{int-csb}  to accommodate simultaneously the new 
CSB result of our study and that of 
a recent cross-section measurement of the isospin forbidden reaction
$dd \rightarrow \alpha \pi^0$ \cite{St03}.

\begin{acknowledgments}
 This work was supported by grants from The Natural Sciences and
 Engineering Research Council of Canada, The National Science 
 Foundation, and The Ohio Supercomputer Center.
 TRIUMF is operated under a grant from the National Research 
 Council of Canada.  
\end{acknowledgments}

\bibliography{e704_prl}

\begin{thebibliography}{9}
\expandafter\ifx\csname natexlab\endcsname\relax\def\natexlab#1{#1}\fi
\expandafter\ifx\csname bibnamefont\endcsname\relax
  \def\bibnamefont#1{#1}\fi
\expandafter\ifx\csname bibfnamefont\endcsname\relax
  \def\bibfnamefont#1{#1}\fi
\expandafter\ifx\csname citenamefont\endcsname\relax
  \def\citenamefont#1{#1}\fi
\expandafter\ifx\csname url\endcsname\relax
  \def\url#1{\texttt{#1}}\fi
\expandafter\ifx\csname urlprefix\endcsname\relax\def\urlprefix{URL }\fi
\providecommand{\bibinfo}[2]{#2}
\providecommand{\eprint}[2][]{\url{#2}}

\bibitem[{\citenamefont{Abegg and {\it et al.}}(1986)}]{tri-csb}
\bibinfo{author}{\bibfnamefont{R.}~\bibnamefont{Abegg}}, 
  \bibinfo{author}{\bibnamefont{{\it et al.}}}, \bibinfo{journal}{Phys.\ Rev.\
  Lett.} \textbf{\bibinfo{volume}{56}}, \bibinfo{pages}{2571}
  (\bibinfo{year}{1986});
\bibinfo{journal}{Phys.\ Rev.  } \textbf{\bibinfo{volume}{D39}}, 
  \bibinfo{pages}{2464} (\bibinfo{year}{1989}).

\bibitem[{\citenamefont{Abegg and {\it et al.}}(1995)}]{tri-new}
\bibinfo{author}{\bibfnamefont{R.}~\bibnamefont{Abegg}}, 
  \bibinfo{author}{\bibnamefont{{\it et al.}}}, \bibinfo{journal}{Phys.\ Rev.
  Let.} \textbf{\bibinfo{volume}{75}}, \bibinfo{pages}{1711}
  (\bibinfo{year}{1995}).

\bibitem[{\citenamefont{{\it et al.}}(1992)}]{iucf-csb}
\bibinfo{author}{\bibfnamefont{L.D.}~\bibnamefont{Knutson}},
  \bibinfo{author}{\bibnamefont{{\it et al.}}}, \bibinfo{journal}{Phys.\ Rev.\ 
  Lett. } \textbf{\bibinfo{volume}{66}}, \bibinfo{pages}{1410} 
  (\bibinfo{year}{1991});
\bibinfo{author}{\bibfnamefont{S.E.}~\bibnamefont{Vigdor}},
  \bibinfo{author}{\bibnamefont{{\it et al.}}}, \bibinfo{journal}{Phys.\ Rev.} 
   \textbf{\bibinfo{volume}{C46}}, \bibinfo{pages}{410} (\bibinfo{year}{1992}).

\bibitem[{\citenamefont{Niskanen}(1999)}]{Ni99}
\bibinfo{author}{\bibfnamefont{J.A.}~\bibnamefont{Niskanen}},
  \bibinfo{journal}{Few-Body Systems} \textbf{\bibinfo{volume}{26}},
  \bibinfo{pages}{214} (\bibinfo{year}{1999}).

\bibitem[{\citenamefont{van Kolck et~al.}(2000)\citenamefont{van Kolck,
  Niskanen, and Miller}}]{vKNM}
\bibinfo{author}{\bibfnamefont{U.}~\bibnamefont{van Kolck}},
  \bibinfo{author}{\bibfnamefont{J.A.}~\bibnamefont{Niskanen}}, 
  \bibnamefont{and}
   \bibinfo{author}{\bibfnamefont{G.A.}~\bibnamefont{Miller}},
  \bibinfo{journal}{Phys.\ Lett.} \textbf{\bibinfo{volume}{B493}},
  \bibinfo{pages}{65} (\bibinfo{year}{2000}).

\bibitem[{\citenamefont{Walden and {\it et al.}}(1999)}]{Wa99}
\bibinfo{author}{\bibfnamefont{P.L.}~\bibnamefont{Walden}}, 
  \bibinfo{author}{\bibnamefont{{\it et al.}}}, \bibinfo{journal}{Nucl.\ Inst.\
  Meth.} \textbf{\bibinfo{volume}{A421}}, \bibinfo{pages}{142}
  (\bibinfo{year}{1999}).

\bibitem[{\citenamefont{Helmer and {\it et al.}}(1987)}]{He87}
\bibinfo{author}{\bibfnamefont{R.}~\bibnamefont{Helmer}},
  \bibinfo{author}{\bibnamefont{{\it et al.}}}, \bibinfo{journal}{Can.\ J.\
  Phys.} \textbf{\bibinfo{volume}{65}}, \bibinfo{pages}{588}
  (\bibinfo{year}{1987}).

\bibitem[{\citenamefont{Hutcheon and {\it et al.}}(2001)}]{Hu01}
\bibinfo{author}{\bibfnamefont{D.A.}~\bibnamefont{Hutcheon}}, 
  \bibinfo{author}{\bibnamefont{{\it et al.}}}, \bibinfo{journal}{Nucl.\ Inst.\
  Meth.} \textbf{\bibinfo{volume}{A459}}, \bibinfo{pages}{448}
  (\bibinfo{year}{2001}).


\bibitem[{\citenamefont{Hutcheon and {\it et al.}}(1991)}]{Hu91}
\bibinfo{author}{\bibfnamefont{D.A.}~\bibnamefont{Hutcheon}}, 
  \bibinfo{author}{\bibnamefont{{\it et al.}}}, \bibinfo{journal}{Nucl.\ Phys.}
  \textbf{\bibinfo{volume}{A535}}, \bibinfo{pages}{618} (\bibinfo{year}{1991}).

\bibitem[{\citenamefont{Auce and {\it et al.}}(1987)}]{deut_cross}
\bibinfo{author}{\bibfnamefont{A.}~\bibnamefont{Auce}},
  \bibinfo{author}{\bibnamefont{{\it et al.}}}, \bibinfo{journal}{Phys.\ Rev.}
  \textbf{\bibinfo{volume}{C53}}, \bibinfo{pages}{2919} (\bibinfo{year}{1987});
\bibinfo{author}{\bibfnamefont{H.}~\bibnamefont{Okamura}},
  \bibinfo{author}{\bibnamefont{{\it et al.}}}, \bibinfo{journal}{Phys.\ Rev.}
  \textbf{\bibinfo{volume}{C58}}, \bibinfo{pages}{2180} (\bibinfo{year}{1998});
\bibinfo{author}{\bibfnamefont{C.}~\bibnamefont{Baumer}},
 \bibinfo{author}{\bibnamefont{{\it et al.}}}, \bibinfo{journal}{Phys.\ Rev.}
 \textbf{\bibinfo{volume}{C63}}, \bibinfo{pages}{037601} (\bibinfo{year}{2001}).

\bibitem[{\citenamefont{Press et~al.}()\citenamefont{Press, Teukolsky,
  Vetterling, and Flannery}}]{num_rec}
\bibinfo{author}{\bibfnamefont{W.}~\bibnamefont{Press}},
  \bibinfo{author}{\bibfnamefont{S.}~\bibnamefont{Teukolsky}},
  \bibinfo{author}{\bibfnamefont{W.}~\bibnamefont{Vetterling}},
  \bibnamefont{and} \bibinfo{author}{\bibfnamefont{B.}~\bibnamefont{Flannery}},
  \eprint{{\it Numerical Recipes,} 1992}.

\bibitem[{\citenamefont{Dumbrajs and {\it et al.}}(1983)}]{Du83}
\bibinfo{author}{\bibfnamefont{O.}~\bibnamefont{Dumbrajs}},
  \bibinfo{author}{\bibnamefont{{\it et al.}}}, \bibinfo{journal}{Nucl.\ Phys.}
  \textbf{\bibinfo{volume}{B216}}, \bibinfo{pages}{277} (\bibinfo{year}{1983}).

\bibitem[{\citenamefont{Coon and {\it et al.}}(1986)}]{Co86}
\bibinfo{author}{\bibfnamefont{S.A.}~\bibnamefont{Coon}},
  \bibinfo{author}{\bibnamefont{{\it et al.}}}, \bibinfo{journal}{Phys.\ Rev.}
  \textbf{\bibinfo{volume}{D34}}, \bibinfo{pages}{2784} (\bibinfo{year}{1986}).

\bibitem[{\citenamefont{Fonseca et~al.}()\citenamefont{Fonseca, G\.{a}rdestig,
Hanhart, Horowitz, Miller, Niskanen, Nogga, and van Kolk}}]{int-csb}
\bibinfo{author}{\bibfnamefont{A.C.}~\bibnamefont{Fonseca}},
  \bibinfo{author}{\bibfnamefont{A.}~\bibnamefont{G\.{a}rdestig}},
  \bibinfo{author}{\bibfnamefont{C.}~\bibnamefont{Hanhart}},
  \bibinfo{author}{\bibfnamefont{C.J..}~\bibnamefont{Horowitz}},
  \bibinfo{author}{\bibfnamefont{G.}~\bibnamefont{Miller}},
  \bibinfo{author}{\bibfnamefont{J.A.}~\bibnamefont{Niskanen}},
  \bibinfo{author}{\bibfnamefont{A.}~\bibnamefont{Nogga}}, \bibnamefont{and}  
  \bibinfo{author}{\bibfnamefont{U.}~\bibnamefont{van Kolk}}, 
  \eprint{private communication}.

\bibitem[{\citenamefont{Stephenson}()}]{St03}
\bibinfo{author}{\bibfnamefont{E.J.}~\bibnamefont{Stephenson}}, \eprint{private
  communication}.

\end{thebibliography}

%\begin{thebibliography}{}
%
%\bibitem{tri-csb} R.\ Abegg, {\it et al.}, Phys.\ Rev.\ Let.\ {\bf 56}, 
%                             2571 (1986);
%      Phys.\ Rev.\ D {\bf 39}, 2464 (1989).
%\bibitem{tri-new} R.\ Abegg, {\it et al.}, Phys.\ Rev.\ Let.\ {\bf 75}, 
%                             1711 (1995);
%          J. Zhao {\it et al.}, Phys. Rev. C 57 (1998) 2126
%\bibitem{iucf-csb} S.E.\ Vigdor, {\it et al.}, Phys.\ Rev.\ C {\bf 46}, 
%                            410 (1992).
%\bibitem{Ni99} J.A.\ Niskanen, Few-Body Systems {\bf 26}, 241 (1999). 
%
%\bibitem{vKNM} U.\ van Kolck, J.A.\ Niskanen, and G.A.\ Miller, 
%        Phys. Lett. {\bf B 493}, 65 (2000).
%
%\end{thebibliography}

\end{document}